%% file: main.tex
\documentclass[10pt,sigconf]{acmart}
\usepackage{subfigure}
\usepackage[inline]{enumitem}
\usepackage[noend,ruled,linesnumbered]{algorithm2e}
\usepackage{graphicx}
\usepackage{float}
\usepackage{bm}
\AtBeginDocument{%
  \providecommand\BibTeX{{%
    \normalfont B\kern-0.5em{\scshape i\kern-0.25em b}\kern-0.8em\TeX}}}

\definecolor{ao(english)}{rgb}{0.0, 0.42, 0.24}

\newcommand{\notes}[1]{\textcolor{ao(english)}{[NOTE: #1]}}
\newcommand{\aunn}[1]{\color{brown}{#1}\color{black}}
\definecolor{dark_green}{RGB}{10,100,70}
\newcommand{\revision}[1]{\color{black}{#1}\color{black}}
\newcommand{\fixme}[1]{\color{blue}{#1}\color{black}}
\newcommand{\shortrev}[1]{\color{black}{#1}\color{black}}

\acmDOI{}

\acmConference[]{}{}

\begin{document}

\title{Adaptive HTAP through Elastic Resource Scheduling}



\author{Aunn Raza}
\email{aunn.raza@epfl.ch}
\affiliation{%
\institution{Ecole Polytechnique F\'ed\'erale de Lausanne}}

\author{Periklis Chrysogelos}
\email{periklis.chrysogelos@epfl.ch}
\affiliation{%
 \institution{Ecole Polytechnique F\'ed\'erale de Lausanne}}

\author{Angelos Christos Anadiotis}
\email{angelos.anadiotis@polytechnique.edu}
\affiliation{\institution{Ecole Polytechnique}}

\author{Anastasia Ailamaki}
\email{anastasia.ailamaki@epfl.ch}
\affiliation{\institution{Ecole Polytechnique F\'ed\'erale de Lausanne}}
\affiliation{\institution{RAW Labs SA}}

\renewcommand{\shortauthors}{Raza, et al.}


\begin{abstract}
Modern Hybrid Transactional/Analytical Processing (HTAP) systems use an integrated data processing engine that performs analytics on fresh data, which are ingested from a transactional engine.
HTAP systems typically consider data freshness at design time, and are optimized for a fixed range of freshness requirements, addressed at a performance cost for either OLTP or OLAP.
The data freshness and the performance requirements of both engines, however, may vary with the workload.

We approach HTAP as a scheduling problem, addressed at runtime through elastic resource management.
We model an HTAP system as a set of three individual engines: an OLTP, an OLAP and a Resource and Data Exchange (RDE) engine.
\shortrev{We devise a scheduling algorithm which traverses the HTAP design spectrum through elastic resource management, to meet the data freshness requirements of the workload. }
We propose an in-memory system design which is non-intrusive to the current state-of-art OLTP and OLAP engines, and we use it to evaluate the performance of our approach.
\shortrev{Our evaluation shows that the performance benefit of our system for OLAP queries increases over time, reaching up to 50\% compared to static schedules for 100 query sequences, while maintaining a small, and controlled, drop in the OLTP throughput.}
\end{abstract}



\settopmatter{printfolios=true}
\settopmatter{printacmref=false}

\maketitle

\input{introduction.tex}

\input{relatedwork.tex}

\input{design.tex}

\input{scheduling-v1.tex}

\input{evaluation.tex}

\input{conclusion.tex}
\balance
\section{acknowledgments}

We would like to thank the reviewers and the shepherd for their valuable feedback.
Angelos Anadiotis was at EPFL until after the initial submission of the paper, and contributed to the revision while at the Ecole Polytechnique.
This work was partially funded by the FNS project "Efficient Real-time Analytics on General-Purpose GPUs" subside no. 200021\_178894/1 and the EU H2020 project SmartDataLake (825041).

\bibliographystyle{ACM-Reference-Format}
\bibliography{sample-base}

\end{document}

%% file: introduction.tex
\section{Introduction}
\label{sec:intro}
Modern business analytics use Hybrid Transactional Analytical Processing (HTAP) systems, where an Online Transaction Processing (OLTP) engine continuously updates the state of the database which serves Online Analytical Processing (OLAP) queries.
HTAP introduces \textit{data freshness} as an additional dimension to analytical data processing.
As data freshness depends on  transactional throughput, an ideal HTAP system provides analytical processing over fresh data without affecting the performance of the transactional engine.
Similarly, analytical processing requires that query response times remain unaffected by data freshness traffic.

\begin{figure}
  \centering
  \includegraphics[width=\columnwidth]{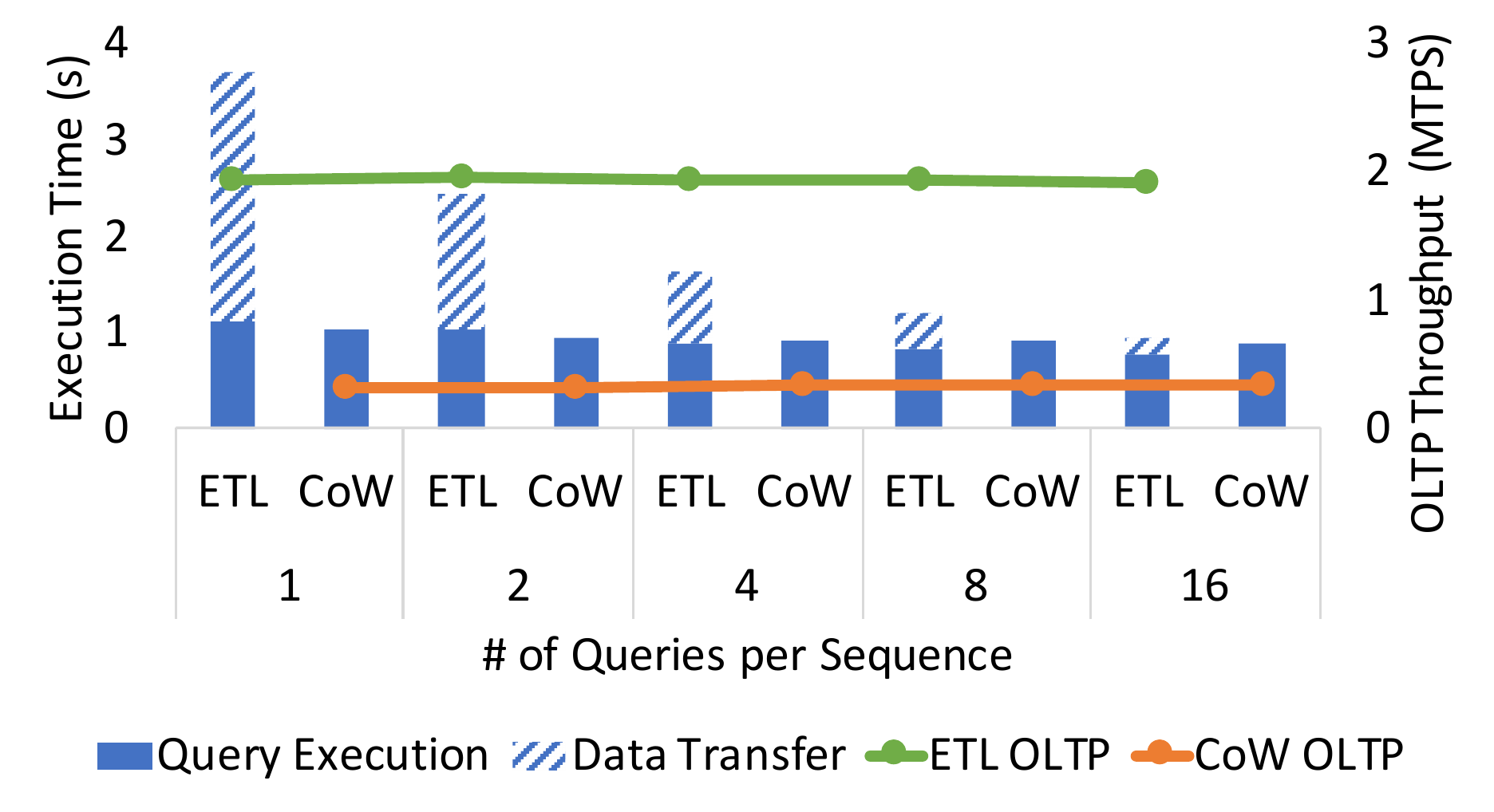} 
  \caption
    {%
      HTAP with ETL and CoW. 
      The engines run on 2 different sockets of a 4-socket server.  
      OLAP executes a batch of 16 aggregate queries and reports the total execution time.
      OLTP executes TPC-C NewOrder transaction, with one warehouse per worker thread.
     }%
     \label{fig:intro-htap-overheads}
\end{figure}

Unfortunately, there is no free lunch.
No matter which mechanism is used to guarantee query execution over fresh data, the performance of either the OLTP or the OLAP part of the HTAP system deteriorates.
\revision{
Figure \ref{fig:intro-htap-overheads} shows examples of the two extremes of the HTAP design spectrum which incur different overhead for each of the two engines.
To demonstrate the performance tradeoff, we execute the same query 16 times and we vary the snapshotting frequency by taking a new snapshot (after OLTP has made updates) at every query, then every two queries, up to sixteen queries, as depicted on the x-axis. 
We measure average query execution time and transactional throughput as per values on the left- and right-hand y-axis, respectively.
}

\revision{
The experiment denoted "ETL" transfers data from the OLTP to the OLAP engine and then executes the analytical query.
Since query execution starts after all updates have been applied to the OLAP engine, the end-to-end query response time is higher.
Nevertheless, the cost is amortized across all queries executed on top of the same snapshot.
The experiment denoted "CoW" relies on hardware-supported Copy-on-Write to allow analytical query execution while running transactions.
When the OLTP engine writes a record while a query is running, operating system first duplicates the affected page, and the write is applied on the copy, thereby ensuring consistency while query execution time remains intact.
Copying pages, however, makes OLTP performance deteriorate.
Without concurrent OLAP queries, OLTP throughput is the same across both experiments (around 2 MTPS).
}

Current state-of-the-art HTAP systems consider data freshness requirements in their design and they can be classified under two categories, based on their storage design.
The first category includes a unified storage, where transactional and analytical workloads are executed concurrently using snapshotting mechanisms.
This category is optimal for analytical workloads where every query accesses further fresh data than the previous one.
The second category includes a decoupled storage, where the OLTP and the OLAP part of the system are executed in isolation and fresh data are transferred from the OLTP to the OLAP part upon request.
This category is optimal for workloads with batches of queries that require the same level of data freshness, or when the data that they need to access is not frequently updated.

However, the amount of fresh data that a query needs to access depends on the workload~\cite{DBLP:journals/pvldb/ChattopadhyayDL19, DBLP:conf/eurosys/CiparGKMSV12}, whereas it can vary at runtime.
For instance, reporting workloads that typically run daily and in batches can be efficiently supported through an ETL process.
On the other hand, real-time statistics workloads require immediate access to fresh data.
Finally, monitoring workloads have queries partly accessing fresh data.
Therefore, a generic HTAP system needs to support workloads accessing different parts of the database which are updated with different frequency, while providing freshness and performance guarantees for both OLTP and OLAP.
Accordingly, the HTAP system has to adapt its design to the workload requirements, at run-time.
In this paper, we refer to the amount of fresh data that a query needs to access, as its \textit{data freshness requirements}. 
%
%
Accordingly, we describe the design and implementation of an HTAP system which supports query workloads with variable data freshness requirements.
Our system adapts to the freshness requirements of the workload, while controlling interference between the transactional and the analytical part, through a scheduling algorithm which distributes compute and memory resources across the OLTP and the OLAP engines.
Our scheduling algorithm is enforced through a system which is non-intrusive to the design of standard in-memory OLTP and OLAP engines and can traverse the HTAP system design spectrum by changing the distribution of resources between the two engines.
Accordingly, we treat HTAP primarily as a resource scheduling problem, where the OLTP and the OLAP engine compete for achieving maximum fresh data locality.
In summary, we make the following contributions:

\begin{itemize}
    \shortrev{\item We approach HTAP as a resource scheduling problem, and we specify a set of states that the system adapts based on the workload requirements for fresh data and the availability of the compute and memory resources that can be exchanged between the engines.}

    \shortrev{\item We show that OLAP performance benefits from getting compute resources of the OLTP engine to bring computation closer to the fresh data, but this benefit is limited by the amount of fresh data accessed and the memory bandwidth.
    In the long run, we demonstrate that the HTAP system converges to a state where all fresh data have to be transferred from the OLTP to the OLAP engine.
    This happens when the size of fresh data dominates the size of the database and, therefore, the data transfer cost is amortized quickly.}
    \shortrev{\item We demonstrate that, compared to static designs, adaptive resource scheduling increases the performance of OLAP query execution up to 50\% for 100 OLAP query sequences while maintaining a small, and controlled, drop in the OLTP throughput.}
\end{itemize}


%% file: relatedwork.tex
\section{Background \& related work}\label{sec:related_work}
\revision{
In this section, we first provide the definition and metric for data freshness that are relevant to the HTAP problem that we study.
Then, we give an overview of existing HTAP approaches and we classify them based on their storage organization, whereas we also present state-of-the-art approaches focusing on the storage layout and in hybrid workload scheduling, which are relevant to our work.
Finally, we classify HTAP workloads based on the expected amount of fresh data which are expected to be accessed during query execution. 
}
\begin{table*}
  \caption{HTAP Design Classification}
  \label{tab:htap_taxonmy}
  \begin{tabular}{lcp{3cm}p{4cm}}
    \toprule
    HTAP Storage & System & Snapshot Mechanism  & Freshness-Perf. Trade-off\\
    \midrule
     & HyPer-Fork~\cite{DBLP:conf/icde/KemperN11}, Caldera~\cite{DBLP:conf/cidr/AppuswamyKPA17} & CoW  & OLTP (CoW) \\
     \textbf{Unified Storage} & HyPer-MVOCC~\cite{DBLP:conf/sigmod/0001MK15}, MemSQL, IBM BLU~\cite{DBLP:journals/pvldb/RamanABCKKLLLLMMPSSSSZ13} & MVCC  & OLAP (version traversal) \\
     & SAP HANA~\cite{DBLP:journals/debu/FarberMLGMRD12} & Delta-Versioning & OLAP (version traversal), OLTP (record chains)  \\
    \midrule
     & BatchDB~\cite{DBLP:conf/sigmod/MakreshanskiGBA17} & Batch-ETL &  OLAP (ETL latency) \\
     \textbf{Decoupled Storage} & Microsoft SQL Server~\cite{DBLP:journals/pvldb/LarsonBHHNP15} & MVCC-Delta &  OLAP (tail-records scan) \\
     & Oracle Dual-format~\cite{DBLP:conf/icde/LahiriCCDGGHHKL15} & Txn Journal \& ETL & OLAP (tail-records scan)  \\
    
    \bottomrule
  \end{tabular}
\end{table*}
\revision{
\subsection{Data Freshness}\label{ref:background_freshness}
In the HTAP design space that we study, we consider two engines, one OLTP and one OLAP, which can be either logically or physically separated.
\shortrev{We suppose that each engine stores data in its private storage, while allowing them to access each other's data through predefined access paths.}
\shortrev{We define as \textit{fresh}, the data resulting from modifications executed by the OLTP engine which are not present in the OLAP private storage when an analytical query arrives. 
Accordingly, the fresh data can be accessed by the OLAP engine either by first copying them to its private storage, or by accessing the OLTP storage directly through access paths exposed by the OLTP engine. 
}
%
\shortrev{Following the definitions in~\cite{DBLP:conf/iqis/BouzeghoubP04}, we measure data freshness with the \textit{freshness-rate metric} which is defined as the rate of tuples that are the same between the private storage of the two engines, over the overall amount of tuples. 
Accordingly, when the two engines share the same data storage, the freshness-rate metric will always be 1.
Instead, when their storage is independent, this metric will generally be less than 1.
}

In HTAP, the analytical queries are executed on top of a data snapshot with freshness-rate metric equal to 1.
This is achieved either by the two engines sharing the same data storage, or by transferring the corresponding delta from the transactional to the analytical storage before the query is executed.
In this paper, we study the performance trade-offs of every approach on the transactional and the analytical engine, and we adapt our system design based on the freshness-rate metric, which we measure at query-level.
}

\revision{\subsection{HTAP engine design}\label{sec:background_design_classification}}
We classify the existing HTAP systems, based on their storage design, in two high level categories: (i) Unified storage, and, (ii) Decoupled storage.
In unified storage, the HTAP system maintains a single consistent snapshot of data for analytical and transactional processing, and isolation between the two engines is achieved through snapshotting.
In decoupled storage, the HTAP system maintains a separate storage for analytical and transactional processing, hence replicating and optimizing data formats while extracting data from the transactional engine, transforming them into the appropriate format, and loading them into the analytical engine.
Table \ref{tab:htap_taxonmy} shows the classification of existing HTAP systems and provides information on the mechanism used for acquiring fresh data snapshots as well as on the trade-offs between performance and data freshness.

\textbf{Unified Storage.} 
The first version of HyPer~\cite{DBLP:conf/icde/KemperN11} relies on partitioned-serial execution for concurrency control among transactions.
Analytical queries are executed on isolated snapshots which are taken lazily upon conflicting access between the transactional and the analytical part of the system.
The snapshot isolation mechanism is based on CoW and Hyper uses UNIX fork to start a new process when an analytical query arrives thereby providing immediate access on the fresh data to the analytical engine.
Caldera~\cite{DBLP:conf/cidr/AppuswamyKPA17} is an HTAP prototype system employing GPUs for analytical query execution, which also relies on CoW using page-shadowing.

The most recent version of Hyper~\cite{DBLP:conf/sigmod/0001MK15} relies on optimistic multi-version concurrency control (MVOCC) to mediate access among transactional and analytical queries.
Similarly, MemSQL and IBM BLU~\cite{DBLP:journals/pvldb/RamanABCKKLLLLMMPSSSSZ13} are commercial HTAP Systems that employ MVCC for snapshot isolation and a tunable option for each table to either store it as a column-major or row-major formats.
SAP HANA~\cite{DBLP:journals/debu/FarberMLGMRD12} is another commercial DBMS system that provides HTAP capabilities using a variant of MVCC-based storage. 
SAP HANA maintains a consistent OLAP-optimized main and OLTP-optimized delta storage which is periodically merged into the main storage. 

HTAP systems with unified storage opt for analytical data freshness using either CoW~\cite{DBLP:conf/damon/MuheKN11} or multi-versioning~\cite{DBLP:journals/pvldb/WuALXP17} for snapshot isolation. 
However, such systems provide data freshness at the cost of performance of either or both analytical and transactional workloads. 
\revision{
Specifically, in CoW, when the OLTP engine updates a record, it has to do a full page copy first, thereby trading transactional for analytical performance.
On the other hand, by using MVCC alone, the OLAP engine has to traverse the versions kept by the OLTP engine, trading analytical for transactional performance due to random memory accesses.
Delta-versioning is a hybrid technique which maintains one OLTP- and one OLAP-optimized snapshot, with second being periodically updated by the first one. 
This approach is fairer since both engines lose performance for accessing each other's snapshot, the OLTP for reading what was recently migrated to OLAP and the OLAP for reading the recently updated data from OLTP.
}
The trade-offs between data freshness and transactional or analytical performance are further analyzed in ~\cite{DBLP:conf/tpctc/PsaroudakisWM00AS14}.


\textbf{Decoupled Storage.}
Traditionally, in data warehousing, the data is extracted from transactional stores, and then transformed and loaded into analytical data stores, during an ETL process. 
Recently, HTAP systems are challenged with more frequent update-propagation mechanisms to provide high data freshness rates. 

BatchDB~\cite{DBLP:conf/sigmod/MakreshanskiGBA17} schedules OLTP and OLAP engines across isolation boundaries, for instance different NUMA nodes, and employs a mini-batching technique to propagate transactional logs to the analytical data store, either periodically or on-demand. 
Oracle's dual-format~\cite{DBLP:conf/icde/LahiriCCDGGHHKL15} maintains OLTP-optimized row-major and OLAP-optimized column-major data in-memory.
Microsoft SQL Server\cite{DBLP:journals/pvldb/LarsonBHHNP15} also maintains two copies of data and propagates data to OLAP storage through an intermediate delta storage to avoid overheads of merging transactionally-hot records, repeatedly. 

\revision{
Decoupled storage has been the conventional way of linking OLTP with OLAP databases.
Batch-ETL provides isolation between the engines, at the cost of OLAP performance for transferring the data from the OLTP snapshot.
However, this cost is amortized by the execution of query batches.
In the MVCC-Delta and the Txn Journal \& ETL approach, the corresponding systems keep a separate snapshot for OLTP and OLAP, and they either transfer periodically the recent versions maintained by the MVCC protocol to the OLAP side, or they use the transactional journal log to do the transfer.
}

\revision{
\textbf{Storage layout.}
NSM and DSM have been the de-facto choices for OLTP and OLAP engines, respectively.
Whereas, PAX~\cite{DBLP:conf/vldb/AilamakiDHS01} addresses hybrid workloads by allowing row and columnar representation in the same disk page.
For in-memory HTAP, systems like SAP HANA~\cite{DBLP:conf/sigmod/Plattner09}~\cite{DBLP:journals/debu/FarberMLGMRD12} and HyPer~\cite{DBLP:conf/icde/KemperN11} have shown that columnar storage~\cite{DBLP:journals/ftdb/AbadiBHIM13} is beneficial for OLAP and only has a negligible performance drop in OLTP. 
Further work on data organization for hybrid workloads~\cite{DBLP:conf/sigmod/DziedzicWDDNS18}~\cite{DBLP:journals/pvldb/AthanassoulisBI19} has shown that with additional optimizations to just static data organization can get orders of magnitude speedup in hybrid workloads.
As the scope of our work is to study and mitigate the effect of performance interference between the OLTP and the OLAP engine, we keep the storage layout constant across the engines and we follow the approach of HANA and HyPer maintaining a columnar layout.
}

\revision{
\textbf{Hybrid workload scheduling.} 
HTAP workload scheduling can be considered as a subclass of the hybrid workload scheduling problem, where the goal is to achieve fairness, while avoiding interference across workloads. 
Works in data-center and cluster scheduling~\cite{DBLP:conf/usenix/IorgulescuAKESN18}~\cite{DBLP:conf/usenix/DelgadoDKZ15} achieve performance isolation between small and large jobs, reducing tail-latencies and head-of-line blocking in a shared cluster. 
Our approach follows the same principles by elastically trading resources between engines through an elastic HTAP scheduler which guarantees that no engine will starve, while allowing every individual engine to optimize their own schedule internally.
However, our focus is on scale-up servers, where the trade-offs are different than scale-out clusters, whereas the metric driving the scheduler decision is the data freshness rate.

Further, run-time resource scaling~\cite{DBLP:conf/sigmod/DasLNK16} and performance isolation~\cite{DBLP:conf/cidr/NarasayyaDSCC13} has been also a well-studied topic in DBMS scheduling.
However, existing work focuses on independent workloads, while in HTAP, workloads have data dependencies due to the data freshness requirements, which lead to concurrent data accesses by the OLTP and the OLAP engine, thus resulting in interference\shortrev{due to implicit (memory bandwidth, CPU caches, hyper-threads) and explicit (locking/latching) resource sharing  caused by OLTP and OLAP's concurrent data accesses.%
}}%

\shortrev{Furthermore, works focusing on scheduling in combination with HTAP traditionally consider the location of the data as input to decide the task placement and, then, the data access method.
For instance, the work of Dziedzic et al.~\cite{DBLP:conf/sigmod/DziedzicWDDNS18} distinguishes and optimizes hybrid workload by adapting the access methods, that are, B+-tree, column-store index or hybrid, on runtime. 
In our approach, the location of the data is one of the outputs of the scheduling algorithm, which decides whether fresh data should be moved from the OLTP to the OLAP engine, or accessed remotely.
Then, based on resource availability, the scheduler decides how to access the data given the amount of fresh data required by every query. 
As we consider an independent storage for each engine, we can physically isolate them and then control interference at the CPU and the memory level by changing the distribution of CPUs across the engines and the data access methods.
}

\subsection{HTAP workload classification}\label{sec:htap_workload_classification}
The amount of fresh data that an OLAP query needs to access, affects the performance of the OLAP and the OLTP engine depending on the design of the HTAP system.
This section provides a classification of HTAP workloads based on their data freshness requirements, following prior works~\cite{DBLP:journals/pvldb/ChattopadhyayDL19, DBLP:conf/eurosys/CiparGKMSV12}. 

\textbf{Short and fresh.} 
This class includes analytical workloads with a high rate of incoming queries. 
The queries are simple, they require fresh data and are mutually independent across the query stream.
Examples of short and fresh analytical workload include dashboard applications where queries access only the latest tail records inserted.
Therefore, the respective queries can be efficiently processed by unified storage engines, since queries access only a part of the data, and they require maximum freshness.

\textbf{Query batches.} 
This class includes recurring and mostly pre-defined queries which are predictable and arrive as a single batch. 
The queries require high, and uniform across all queries in the batch, data freshness, since the results reflect the state of the database until a certain point in time.
A standard example is the reporting queries generated periodically for the state of the sales of a business unit or the state of an area which is under electronic access control.
Therefore, the respective queries can be efficiently processed by decoupled storage engines, since queries need to access exactly the same data, which have to be as fresh as possible.

\textbf{Ad-hoc queries.}
This class includes dynamic queries requiring access to fresh data, whereas they can combine both fresh and old data.
Queries for predictions and forecasting require such combination of historical and fresh data, and they are typical examples falling into this category.
Therefore, the performance of an HTAP system is in this case by the amount of fresh data that each query will access.
Accordingly, in case a query needs to access mostly fresh data, a unified storage engine is the most appropriate choice.
Instead, queries that access mostly historical data, would benefit mostly by a decoupled storage design.
However, this information is only known at run-time, and therefore the system has to adapt to the size of fresh data accessed by each quer.y

%% file: design.tex
\section{System design \& implementation}
\begin{figure*}[ht]
    \centering
    \includegraphics[width=\textwidth]{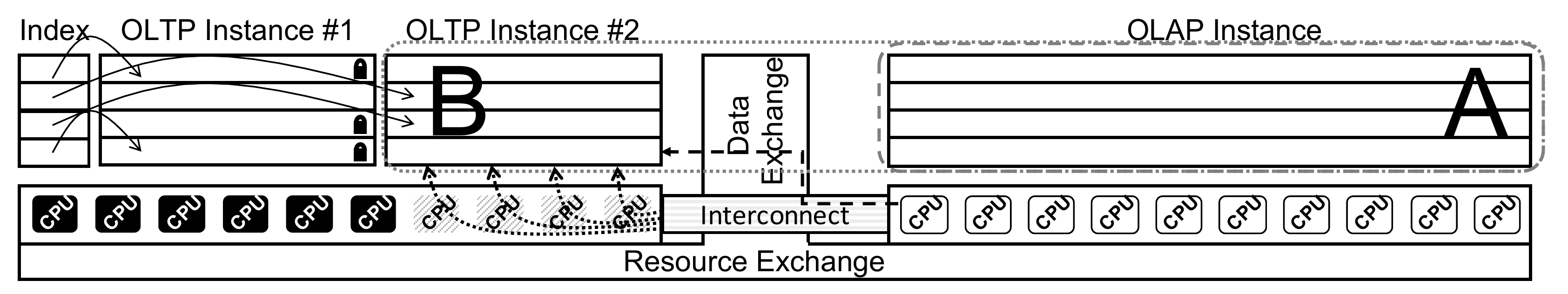}
    \caption{System architecture}
    \label{fig:design_overall}
\end{figure*}
This Section describes the design of our HTAP system. 
As we consider two independent engines that can work together through a thin, scheduling layer, we split our description into four parts.
First, we provide an overview of the overall system to give the big picture and link with the design choices of each individual engine.
Then, we provide the technical description of the OLTP and the OLAP engines and show that our design is non-intrusive to the standard design of such in-memory engines.
Finally, we explain how data and resources are exchanged between the OLTP and the OLAP engine through the Resource and Data Exchange engine.

\subsection{Overview}
Hybrid Transactional/Analytical Processing addresses workloads where analytical queries process fresh data.
In the presence of both the OLTP and the OLAP engine, data freshness is guaranteed either by sharing the storage of the OLTP engine or by copying a part of it to the OLAP engine.
There are two important challenges associated with achieving query execution over fresh data:
(i) explicit resource sharing causes interference both at the software (extensive copy-on-write or non-contiguous scans) and at the hardware level (sharing CPUs and memory bus), and, 
(ii) data copying imposes an increased latency which may not be tolerable by the workload.
We address these challenges through an elastic system design which adapts to the workload requirements.

Figure~\ref{fig:design_overall} depicts the architecture of our HTAP system.
Our system achieves resource isolation and sharing on demand, with respect to the performance requirements of each engine. 
\shortrev{OLTP and OLAP engines execute the workload independently, and have one-way dependency: only the OLAP engine reads  fresh-data from the OLTP.
As sessions do not span across the two engines, each engine maintains its own request queue and executes the requests independently. 
The RDE engine assigns resources (CPU and memory) and data to the engines.
}
Following the common approach in cloud computing, we assume that CPU and memory resources are split in two sets: the first is exclusively given to each engine, and the second can be traded between them.
The distribution of resources between the engines is decided by the RDE engine.
By introducing RDE as an integration layer, we achieve an adaptive HTAP system design, which is minimally intrusive to the design of existing OLTP and OLAP engines.

\subsection{OLTP engine}
The OLTP engine is depicted on the left-hand side of Figure~\ref{fig:design_overall}.
Following the standard in-memory OLTP system design, our engine includes a Storage Manager (SM), a Transaction Manager (TM) and a Worker pool Manager (WM).

\textbf{Storage Manager.} 
The SM stores data entirely in the main memory in columnar format. 
Following an approach similar to Twin Blocks~\cite{DBLP:conf/sigmod/CaoSSYDGW11} and Twin Tuples~\cite{DBLP:conf/damon/MuheKN11}, the SM maintains two \textit{instances} of the data, in addition to a multi-versioned storage.
The difference from existing approaches is that each instance keeps data in a columnar layout, to allow the OLAP engine to perform fast scans over the data without having to traverse specific versions.
At every time point, only one of the two instances is active.
The OLTP engine also maintains an index, implemented using cuckoo hashing~\cite{DBLP:journals/jal/PaghR04}.
The index always points to the last updated record in either of the two instances.
Therefore, even though the inactive instance will remain out-of-date, upon switch, the data will always come from the newest version.
The OLTP engine also maintains a delta storage to allow transactions to traverse older versions of the objects in Newest-to-Oldest ordering~\cite{DBLP:journals/pvldb/WuALXP17}, following the standard multi-versioned concurrency control (MVCC) process.
By maintaining two explicit instances, we split the storage into a part which is contiguous and another one where data are stored in random order.
This way, in case there are at least two versions of the data, we do not impose any further requirements on memory size.
The active instance of the SM is switched upon request.
The SM provides an API to switch the instance, which returns the starting address of the inactive instance when no active OLTP worker thread is using it any more. 
\revision{The SM maintains instance statistics per column, which are the number of records at the time of switch, a flag indicating if the column contains updated-tuples and the epoch number. }
The SM also maintains an \textit{update indication bit} for each record, which is set when the record gets updated.
Access to the update indication bits is synchronized using atomic operations, as they can be accessed by different parts of the system.

\textbf{Transaction Manager.} Together with the SM, the TM provides transactional access to the database records by relying on a two-phase locking (MV2PL~\cite{DBLP:journals/pvldb/WuALXP17}) concurrency control protocol with deadlock avoidance~\cite{DBLP:journals/pvldb/AppuswamyAPIA17} \revision{and transactional isolation level of snapshot isolation~\cite{DBLP:conf/sigmod/BerensonBGMOO95}}. 
When a transaction tries to access a record, it brings from the index its most recent value.
This value may either be on the currently active instance or on the inactive one.
When a transaction starts, it requests the SM for the starting address of the column which is on the currently active instance, and performs all the operations there, sets the update indication bit for that record, and then updates the index, if needed.
Every update is directly placed on the active instance upon transaction commit, and the older version is pushed to the versioned storage.
This provides a significant speedup to the OLAP query execution since it allows sequential scans.
Inserts are pushed to both instances, but they are made available through the inactive instance by the Storage Manager only after a switch.

\textbf{Worker Manager.} The OLTP engine uses one hardware thread per transaction.
The WM keeps a worker pool of active threads.
We set each thread to first generate a transaction and then execute it, simulating a full transaction queue.
The WM exposes an API to set the number of active worker threads and their CPU affinities, thus enabling the OLTP engine can elastically scale up and down upon request.

\subsection{OLAP engine}
\label{sec:design-olap}
The OLAP engine is based on Proteus~\cite{DBLP:journals/pvldb/KarpathiotakisA16,DBLP:journals/pvldb/ChrysogelosKAA19}, a parallel, NUMA-aware, query execution engine with support for CPU-GPU execution~\cite{DBLP:journals/pvldb/ChrysogelosKAA19}.
The engine uses code generation to specialize the executed code to each query~\cite{DBLP:journals/pvldb/Neumann11}, compute device~\cite{DBLP:journals/pvldb/ChrysogelosKAA19} and access path~\cite{DBLP:journals/pvldb/KarpathiotakisA16}.
Our design considers CPU as the only compute device available, leaving hardware accelerators as future work.
Following the description of the OLTP engine, we describe how the Storage Manager (SM) and Query Executor (QE) of the OLAP engine, work.
The OLAP engine also includes a Worker Manager (WM), which works in a similar way to the WM of the OLTP engine, and therefore, we omit its description to avoid redundancy.

\textbf{Storage Manager.}
The SM considers that data are stored in the main-memory of a single server and it is agnostic of their format and layout.
The data access paths are decided by input plugins which specify how tuples are accessed and interpreted, following the schema of the database and the current data format.
The input plugins further encapsulate different data access methods.
In our HTAP setting, we use two access methods.
The first method considers that data are stored in the same contiguous memory area.
The second method considers that data are partitioned in several (contiguous) memory areas, and it is useful when we need to access only the fresh data from the OLTP storage and the rest from the OLAP storage.
The SM accepts as input a pointer to the memory areas where the data are stored at execution time, and it does not load any data beforehand.

\textbf{Query Executor.}
The OLAP engine generates just-in-time specialized code for each query, using the appropriate plugins to access the data.
The query plan is translated into pipelines which execute sequences of operations on input tuples without materializing intermediate results.
Each pipeline is transformed into code which is compiled and optimized for the current hardware architecture.
By default, the pipelines process one block of tuples at a time.
During code generation, the plugins specialize the different access methods, based on the location of the data. 

The OLAP engine parallelizes query execution by routing blocks between different pipelines that execute concurrently.
Each pipeline is assigned to a worker which is affinitized to a CPU core.
Based on the placement of the data, the OLAP engine balances the load across worker threads using protocols (hash-based, load-aware, locality-aware and combinations).
By default, the OLAP engine uses locality-and-load-aware policies, and schedules blocks to pipelines which are executed locally to the data, if possible. 
When a data block is remote to its worker thread, the OLAP engine either accesses the data over the interconnect, or it prefetches the block to the local CPU socket, while overlapping the data transfer with the execution over other blocks.
The engine chooses the best strategy based on the availability of resources.


\subsection{RDE engine}\label{sec:rde_engine}
The Resource and Data Exchange engine is the integration layer between the OLTP and the OLAP engines and supports the operations required for HTAP.
Fundamentally, there are two ways to achieve HTAP and they require two different design choices.
The first way is to have the two engines running in isolation, which we assume to be at the socket boundary in a single, scale-up server.
HTAP is thus supported by transferring fresh data from the OLTP to the OLAP before executing a query.
The second way is to have the two engines sharing their resources, effectively having a single HTAP system with transactional and analytical processing capabilities.
Our HTAP system design relies on elastic resource management to traverse the design space between the above two approaches, through the RDE engine.

In the following, we first provide fundamentals for the RDE engine design, then we describe the discrete states the system marking them as $S_1$ (Co-located OLTP and OLAP), $S_2$ (Isolated OLTP and OLAP), and $S_3$ (Hybrid), and finally we explain how the RDE engine elastically migrates between states.
The decision on which state to move at each point in time is taken by the system scheduler and enforced by the RDE engine.
Our system design is independent of the scheduling algorithm, and therefore, we leave the description of the latter to the next Section and here we focus only on the way our system can adapt to different configurations that are required by different HTAP workloads.


The RDE engine is the owner of memory and CPU resources and distributes them to the OLTP and the OLAP engines.
\revision{
Still, each engine has its own internal scheduler that decides which resources to use based on the workload.
For instance, the scheduler of our OLAP engine makes NUMA-aware decisions on the use of multiple nodes for each query operator.
Resources that are not accepted by an engine are returned to the RDE and offered to the other engine.
}

\textbf{OLTP active instance switching.}\phantomsection\label{sec:switchtime}
The OLTP engine provides \revision{consistent and }fresh data with\revision{ with snapshot isolation guarantees}.
To avoid interference with transaction execution every time the OLAP engine needs access to fresh data, the RDE engine instructs the OLTP engine to switch its active instance.
However, this creates a freshness-level inconsistency between the two instances, which can increase with time.
For instance, if some records get updated every two OLTP instance switches, then they will be fresh in one instance, leaving the other one behind.
\revision{
Upon the instance switch, the RDE engine checks for updates through a hierarchical update-presence flag in the order of schema, relations, and columns. In the presence of updates, the RDE engine traverses the update indication bit and for the records that are updated, it copies them to the other instance, in case they have not been updated there as well by that time.
As the number of inserts and updates corresponds to the data freshness rate, the RDE also maintains these statistics which are provided to the scheduler.
}
With careful engineering, this process has a negligible effect in performance, and it is followed every time we refer in the following to switching the active instance of the OLTP engine.
It takes around 10ms to sync around 1 million \revision{modified tuples in a database of over 1.8 billion records, while executing TPC-C NewOrder transaction. The size of database corresponds to TPC-H scale-factor 300.}

\textbf{Co-located OLTP and OLAP $\boldsymbol{[S_1]}$.}
In this state, the two engines share the memory and the CPUs of all the sockets.
The CPUs of each socket are distributed to the engines following the decision made by the scheduler.
When an OLAP query arrives, the RDE engine instructs the OLTP engine to switch its active instance.
The OLTP engine returns the pointer of its inactive instance, which is then used by the OLAP engine to execute the query.
This way, the two engines interfere at the hardware, but not at the software level, since the OLAP query is executed on a part of the memory which is not used by the OLTP engine.
The OLTP engine continues transaction execution on its own instance of the data.

An example of the distribution of resources is depicted in Figure~\ref{fig:design_overall}, on the left-hand socket, where black colored CPUs are used for OLTP and the stripped ones are used for OLAP query execution.
The black colored CPUs access only OLTP Instance \#1, whereas the stripped ones access only OLTP Instance \#2.
The RDE engine changes the ownership of the respective part of the memory between the two engines.
In this setting, the OLTP engine has access to both the instances, since the index may point to data stored in the instance used by the OLAP engine.
As both engines only read data from the second instance, there is no conflicting operation involved and therefore no need for synchronization.

\revision{
\textbf{Related systems to $\boldsymbol{[S_1]}$.}
This state represents the class of systems which employ co-location of compute and storage resources, like SAP HANA~\cite{DBLP:journals/debu/FarberMLGMRD12} and HyPer-MVOCC~\cite{DBLP:conf/sigmod/0001MK15}. 
Co-location of hybrid workloads also represents CoW based systems, since the OLAP engine gets fresh snapshots instantly while OLTP still proceeds on secondary data instance.
Our design avoids the CoW overheads for OLTP while providing OLAP access to data stored in columnar layout.
}

\textbf{Isolated OLTP and OLAP $\boldsymbol{[S_2]}$.}
In this case, the two engines run in the highest isolation level with minimal interference.
Each engine receives resources at the granularity of a CPU socket, following the decision made by the scheduler.
When an OLAP query arrives, the RDE engine instructs the OLTP engine to switch its active instance.
After the OLTP engine returns the pointer to the inactive instance, the RDE engine transfers the data that have been inserted and the data that have been updated since the last time that the instance of the OLAP engine was updated.
Updated data are recognized by the update indication bit, which is set by the OLTP engine.
For each record transferred to the OLAP engine, the RDE engine clears the corresponding bit.
Data transfers are overlapped with OLTP instance synchronization to avoid re-reading the same records.
Even though CPU-level isolation is achieved throughout the whole query execution time, there is interference at the memory level when reading the data from the socket where the OLTP engine is executed.
This interference is limited by the interconnect bandwidth, which is typically several times smaller than the memory bandwidth, whereas the OLTP engine in any case does not fully utilize memory bandwidth due to random memory accesses.

An example of the distribution of resources to the engines is shown in Figure~\ref{fig:design_overall}, considering that the OLTP engine occupies the full left-hand socket and the OLAP engine the full right-hand side socket with the long dashed line indicating the memory isolation boundary.
Supposing that OLTP Instance\#2 was active when the query arrived, after the switch, its data are transferred by the RDE engine to OLAP instance through the socket interconnect.
Given that analytical query execution cannot start before all data have been transferred, the RDE engine uses resources of OLAP to transfer the data.
Therefore, data transfer time is accounted to the overall query execution time, as there is no benefit in hurting the performance of the OLTP engine at this point.

\revision{
\textbf{Related systems to $\boldsymbol{[S_2]}$.}
This state represents the class of systems with decoupled storage and full compute isolation, like BatchDB~\cite{DBLP:conf/sigmod/MakreshanskiGBA17} and traditional data-warehouse solutions, where a periodic ETL is performed from OLTP to OLAP engines.
These systems isolate OLTP and OLAP workloads across hardware boundaries, NUMA or machines, and provide software and hardware level  isolation. 
}

\textbf{Hybrid OLTP and OLAP $\boldsymbol{[S_3]}$.}
In this case, the two engines share memory and, if requested, CPU resources.
The key aspects of the hybrid approach are: (i) the OLAP engine accesses only the fresh data that it needs for a specific query, and, (ii) the OLAP engine accesses fresh data either through the interconnect or directly from the socket where the OLTP engine is executed, by taking some CPUs from the OLTP engine.
Similarly to the previous states, when an OLAP query arrives, the RDE engine instructs the OLTP to switch its active instance and passes the pointer to the inactive instance to the OLAP engine.
Then, the OLAP engine has two options: it either accesses the fresh data through the interconnect, like in the isolated state, or it gets some CPUs on the socket of the OLTP engine and accesses fresh in full memory bandwidth from these CPUs.
The scheduler decides which of the two options to be used.
The interference caused by the hybrid approach at the memory level is bounded at the lower side by the interconnect bandwidth, whereas at the CPU level is bounded at the higher side by the number of CPUs allowed to be passed from the OLTP to the OLAP engine, by the database administrator.

An example of the hybrid approach can be reconstructed from Figure~\ref{fig:design_overall}, by considering cases A and B referring to the first and the second option, respectively.
In case A, the OLAP engine uses its own socket and accesses fresh data from the interconnect, as the long-dashed-line arrow indicates.
In case B, the OLAP engine uses the stripped CPUs from the socket of the OLTP engine, and after performing some operations on that socket, they send the data back to the main OLAP socket, as the short-dashed-line arrow indicates.
Case B is particularly useful for query operators with a big reduction factor (e.g., an aggregation like a count), which would stress the interconnect if case A would be followed.

\revision{
\textbf{Related systems to $\boldsymbol{[S_3]}$.}
This state represents the class of systems employing hybrid data access techniques, like tail OLTP record scan for OLAP, which is equivalent to accessing fresh data from the OLTP inactive instance in our case, in addition to scanning OLAP-local storage. 
Representative systems includes Microsoft SQL Server~\cite{DBLP:journals/pvldb/LarsonBHHNP15}, and Oracle Dual-format~\cite{DBLP:conf/icde/LahiriCCDGGHHKL15}.
Further, state $S_3$ follows the design of elastic resource allocation in cloud systems to distribute resources between the two engines at runtime.
}

%% file: scheduling-v1.tex
\section{Adaptive HTAP scheduling}
The elasticity at the system design level is driven by a scheduler which decides how to distribute resources to the OLTP and the OLAP engine.
The main parameter considered by the scheduler is the \revision{data freshness rate for each query}.
\shortrev{Accordingly, the scheduler selects a system state and the OLAP engine adapts its resource allocation and data access methods to provide maximum data freshness for each analytical query. }
As execution ranges across three states: co-location, isolation, and hybrid, the scheduler requests the system to migrate to a state by changing the size of the worker pool of each engine and the affinity of the worker threads.
Accordingly, workload execution adapts to the resources that are made available to the engines every time.

\subsection{System model}
We model the HTAP system as a set of memory and computing resources and we assign them to the engines, which use them internally to optimize their execution.
We rely on the scan operators of the analytical query plan to find the data that the query will access and we extract the fraction of fresh data.
Further, we assume that data are stored in a columnar layout and that there no additional overheads to scanning other than any potential NUMA effects that are caused by using CPUs that are remote to the data.
Observe that NUMA overheads are fully controllable by the scheduler, since it distributes the resources to the engines.
Finally, we assume that the database administrator can set restrictions on the amount of resources that can be revoked from the OLTP and the OLAP engine, to abide with performance guarantees and control interference between the engines.

\shortrev{
We do not consider any optimizations through indices or related data structures and we assume that the system will perform a full scan of each column.
In the presence of such optimizations, the HTAP system will still need to schedule the maintenance of the associated data structures considering the data updates.
Even though we expect that our scheduler will still be applicable, we leave this as future work which can rely on our current findings.
}

The main overheads that we consider in the execution are the remote memory accesses, since they can affect significantly execution times, especially in OLAP engines that have to process big amounts of data.
Given that state-of-the-art analytical query engines can saturate the memory bandwidth while scanning the data, we can quantify the overhead for remote vs local memory access to be equal to the difference in bandwidth between the main memory bus and the CPU interconnect.
For OLTP engines, which are characterized by random data accesses and therefore use only a part of the memory bandwidth, we assume that the overhead for remote data access is less than the analytical ones.
Given that the profile of the transactions typically does not change over time, we assume that the scheduler can easily learn and experimentally quantify this overhead and adjust it at run-time, if the profile of the transactional workload changes.
Finally, both engines can scale as they use more CPU cores even from a socket which is remote to the data, despite the interconnect bottleneck, which leads to lower performance.

The scheduler makes decisions in two levels. 
The first level is the state selection.
The scheduler selects the state that optimizes access locality to the fresh data required by the query.
The second level is the resource distribution.
Memory-wise, in the co-located state, the OLTP and the OLAP engine take one of the OLTP instances each.
In the isolated case, the OLTP and the OLAP keep their own instances.
In the hybrid case, the OLTP keeps one of its active instances and shares the second with OLAP.
Computing-wise, in the co-located state, the OLTP and the OLAP share at maximum all CPU sockets.
In the isolated case, there is no change in the CPU distribution.
In the hybrid case, the OLAP engine may use a number of CPU cores that primarily belonged to the OLTP engine and interfere with its execution.
The thresholds on the number of CPU cores to exchange between the two engines are set by the database administrator, as they affect workload execution in both engines, and therefore resource allocation has to remain compliant to any performance guarantees.

\subsection{Elastic resource scheduling}\label{sec:elastic_resource_scheduling}

The adaptivity of our HTAP system is achieved through fine-grained, elastic resource scheduling, which allows the system to migrate across different states.
\shortrev{Elasticity allows the RDE engine to provide different data-access paths to the OLAP engine to achieve maximum data freshness, while having controlled interference in OLTP performance. }
This section first describes how state migration is achieved by setting the CPU and memory resources of the OLTP and the OLAP engine.
Then, it describes how data freshness drives the decision for migrating across states.
\input{algorithms/migrate_state_s1.tex}

Algorithm~\ref{algo:state_migration} describes the steps required to migrate the system to one of its states.
First, we define two thresholds for the minimum computing resources that have to be given to the OLTP engine, at socket and CPU granularity.
The thresholds are useful in two states: (i) the co-located one, where the OLTP and the OLAP share some or all the sockets of the server, and, (ii) the hybrid-elastic one, where the OLAP engine uses a set of CPUs that belonged to the OLTP engine.
Then, we provide each state migration as a separate function.
For each function, the scheduler only assigns resources; their enforcement is performed by the RDE engine.

\revision{
The \texttt{MigrateState$S_1$} function assigns the number of CPUs defined by the corresponding threshold to the OLTP engine, and the rest to the OLAP engine.
Then, it switches the active instance of the OLTP engine, and sets the OLAP engine to read data from the other, now inactive, OLTP instance.
The \texttt{MigrateState$S_2$} function distributes the system CPU sockets to the OLTP and the OLAP engine according to the policy set by the database administrator.
For instance, in the case of a uniform policy, OLTP and OLAP engines will get half of the available sockets.
Then, the scheduler marks the memory for the ETL and requests OLAP to use its local instance.
As ETL is performed by the RDE engine, the latter uses OLAP compute resources while transferring the data, given that OLAP cannot execute the query before all data have been copied.
The ETL process triggered copies only the delta between the two instances, which is calculated by the RDE engine based on the update indication bits set by the OLTP engine.
The \texttt{MigrateState$S_3$} function is similar to the other two, depending on the thresholds set for OLTP resource allocations.
In the \texttt{ISOLATED} mode, we set the compute resources to socket-level isolation, and the OLAP engine accesses records from the OLTP engine over the interconnect.
In the \texttt{NON-ISOLATED} mode, the OLAP engine is co-located in some sockets with the OLTP engine by getting some of its CPU cores, thus favoring OLAP over OLTP performance.
}

\input{algorithms/freshness_scheduling.tex}
\revision{
Algorithm~\ref{algo:freshness_scheduling} describes the scheduling strategy for migrating the system across different states.
The decisions of the algorithm are based on the freshness-rate metric for every query.
Recall that the freshness-rate metric in our HTAP system is the rate of tuples that are the same in the OLAP and the currently active OLTP instance when the query arrives.
\shortrev{Algorithm~\ref{algo:freshness_scheduling} calculates the freshness-rate metric only for the columns which will be accessed by every query.
}
The scheduler retrieves from the RDE engine the amount of fresh data that the OLAP engine needs to fetch from the OLTP instance to satisfy the current query $N_{fq}$ with freshness-rate 1 and the amount of fresh data to update the whole OLAP instance $N_{ft}$.
The parameter $\alpha$ is defined within $\left[0,1\right]$ and as it decreases, the scheduler prefers to do ETL by migrating to $S_2$.
\shortrev{If ETL is not preferred due to the amount of fresh data, the scheduler will check whether elasticity is allowed, denoted with the flag $F_{el}$. }
If it is not allowed, then it will instruct the OLAP engine to read the data needed for the query remotely from the OLTP instance, after migrating to state $S_3-ISOLATED$ $(S_3-IS)$.
If elasticity is allowed, then the system will either migrate to the state $S_3-NON-ISOLATED$ $(S_3-NI)$ or to $S_1$, depending on the performance requirements of the OLTP engine.
Therefore, the decision on the elasticity mode $M_{el}$ is based on the service level agreement for the OLTP engine.
The number of CPU cores to be passed from the OLTP to the OLAP engine is subject to the workload and the OLTP performance requirements. 
We conduct a sensitivity analysis to explain the numbers that we are considering for our evaluation in Section~\ref{sec:eval_sensitivity_analysis}.

Given that the performance of an HTAP system is determined by the performance of both the OLTP and the OLAP engine, Algorithm~\ref{algo:freshness_scheduling} is a heuristic which tries to optimize the performance of OLAP given the restrictions of the OLTP engine.
For this reason, it first favors for OLAP to take compute resources from OLTP ($S_3-NI$), then to trade them with the OLTP ($S_1$) and finally to just do remote access ($S_3-IS$).
In all cases, when there is enough fresh data, as defined by $\alpha$, the algorithm migrates to $S_2$ to keep the OLAP instance fresh and provide data locality for future queries.

\shortrev{\textbf{Query Batch.} }We consider as batch, a set of queries that are executed over the same data snapshot with the same freshness rate.
Thus, the execution of the batch depends only on the OLAP engine and it is orthogonal to the scheduler.
As the number of queries is increased, so does the probability of them accessing all the fresh data of the OLTP instance, making $N_{fq}$ to approach $N_{ft}$ and leading the scheduler to migrate to $S_2$.
This also applies for individual queries, where the scheduler is expected to initially trigger states $S_1$ and $S_3$ which do not update the OLAP instance, but at some point the rate of fresh data per query to the overall amount of fresh data will approach 1, eventually migrating to $S_2$.

\shortrev{\textbf{ETL Sensitivity.} 
ETL sensitivity is given by the parameter $\alpha$ in Algorithm~\ref{algo:freshness_scheduling} and it represents the threshold for copying the fresh data from the OLTP to the OLAP instance.
Small values of $\alpha$ increase the sensitivity of the scheduler into performing an ETL by migrating to state $S_2$.
This is beneficial for workloads where every query is expected to touch the same attributes as the previous ones, or workloads where only a small fraction of the data gets updated.
Instead, big values of $\alpha$ are beneficial for workloads where every query is expected to access a small subset of the updated data.
}

\shortrev{\textbf{Elasticity and Interference.} 
Elasticity introduces interference between the OLTP and the OLAP engine.
Bandwidth-intensive OLAP can starve OLTP with only a few hardware threads by consuming memory bandwidth apart from the compute resources.
Limits in the use of CPUs and memory bandwidth can be set by using hardware tools of server-grade CPUs~\cite{intel_rdt} or software-based solutions~\cite{NarasayyaDSCC13, DBLP:conf/isca/LoCGRK15}.
In~\cite{our_techreport}, we analyze the effect of elasticity on interference between the workloads. 
To better utilize the hardware resources, real-time performance monitoring can be employed~\cite{DBLP:conf/cloud/FarleyJVRBS12, DBLP:conf/osdi/GrandlKRAK16, DBLP:conf/osdi/MahajanCA018}, as they will allow the scheduler to distribute resources between the engines until a certain performance degradation threshold.
}

}

%% file: algorithms/migrate_state_s1.tex
\begin{algorithm}

\SetKwProg{Fn}{}{}{}
\SetAlgoLined
\DontPrintSemicolon
\KwData{OLTPSockThres = Minimum OLTP Sockets}
\KwData{OLTPCpuThres = Minimum OLTP CPUs/Socket}
\SetKwFunction{MigrateStateSone}{MigrateState$S_1$}%
\SetKwFunction{MigrateStateStwo}{MigrateState$S_2$}%
\SetKwFunction{MigrateStateSthree}{MigrateState$S_3$}%
\AlgoDisplayBlockMarkers\SetAlgoBlockMarkers{}{}%
 \Fn{\MigrateStateSone{}}{
    \For {s in Server.CpuSockets} {
        \While {OLTP.cpuCnt < OLTPCPUThres[s]} {
            OLTP.addCPU(s.nextCPU)
        }
        OLAP.addCPU(s.nextCPU)
    }
    OLAP.setMem(OLTP.switchInstance())
}
\BlankLine
 \Fn{\MigrateStateStwo{}}{
    \For {s in Server.CpuSockets} {
        \If{OLTP.socketCnt < OLTPSockThres}{
            OLTP.addSocket(s)
        }
        \Else{
            OLAP.addSocket(s)
        }
    }
    OLAP.etl(OLTP.switchInstance())
    OLAP.setMem(OLAP.localInstance())
}
\BlankLine
 \Fn{\MigrateStateSthree{mode}}{
    \If{mode == ISOLATED}{
        \If{OLTP.socketCnt < OLTPSockThres}{
            OLTP.addSocket(s)
        }
        \Else{
            OLAP.addSocket(s)
        }
        OLAP.setMem(OLTP.switchInstance())
    }
    \Else {
        \For {s in Server.CpuSockets} {
            \While {OLTP.cpuCnt < OLTPCPUThres[s]} {
                OLTP.addCPU(s.nextCPU)
            }
            OLAP.addCPU(s.nextCPU)
        }
        OLAP.setMem(OLTP.switchInstance())
    }
}

\caption{State Migration}
\label{algo:state_migration}
\end{algorithm}

%% file: algorithms/freshness_scheduling.tex
\begin{algorithm}

\SetKwProg{Fn}{}{}{}
\SetAlgoLined
\DontPrintSemicolon
\KwData{$F_{el}$ = Elasticity availability flag}
\KwData{$M_{el}$ = Elasticity mode: \{Hybrid, Co-location\}}
\KwData{\revision{$N_{fq}$ = Amount of fresh data in query}}
\KwData{\revision{$N_{ft}$ = Amount of fresh data in database}}
\KwData{\revision{$\alpha$ = ETL sensitivity}}
\SetKwFunction{ResourceSchedule}{ResourceSchedule}%
\SetKwFunction{MigrateStateSone}{MigrateState$S_1$}%
\SetKwFunction{MigrateStateStwo}{MigrateState$S_2$}%
\SetKwFunction{MigrateStateSthree}{MigrateState$S_3$}%
\AlgoDisplayBlockMarkers\SetAlgoBlockMarkers{}{}%
 \Fn{\ResourceSchedule{}}{
    \If {\revision{$N_{fq} < \alpha N_{ft}$} \textbf{AND} !QueryBatch} {
        \If{$!F_{el}$}{
            \MigrateStateSthree{ISOLATED}
        }
        \ElseIf{$M_{el} == HYBRID$}{
            \MigrateStateSthree{NON-ISOLATED}
        }
        \Else{
            \MigrateStateSone{}
        }
    }
    \Else {
        \MigrateStateStwo{}
    }
}
\caption{Freshness-driven resource scheduling}
\label{algo:freshness_scheduling}
\end{algorithm}



            
        

%% file: evaluation.tex
\section{Experimental Evaluation}\label{sec:expeval}

This section includes the results of our experimental evaluation.
First, we describe the hardware that we used to execute our experiments, some essential details of our software, and finally the benchmark that we used to derive our workload.
\revision{
Then, we present the results of our sensitivity analysis focusing on specific queries, to show the benefits of every state of our system.
Based on the sensitivity analysis, we tune our scheduler and we evaluate the performance of every state of the system under different HTAP queries.}
Finally, we derive a query mix and we evaluate the adaptivity properties of our scheduler throughout the workload execution.

To the best of our knowledge, there is no single performance metric, which is defined for HTAP workload execution.
Accordingly, in our evaluation, we focus on the way that our system selects the optimal configuration, based on the level of interference which is allowed between the engines.

\begin{figure*}
  \centering
  \begin{minipage}[t]{.33\textwidth}
  \centering
  \subfigure[OLTP/OLAP performance at State $S_1$]{\centering\includegraphics[width=\textwidth]{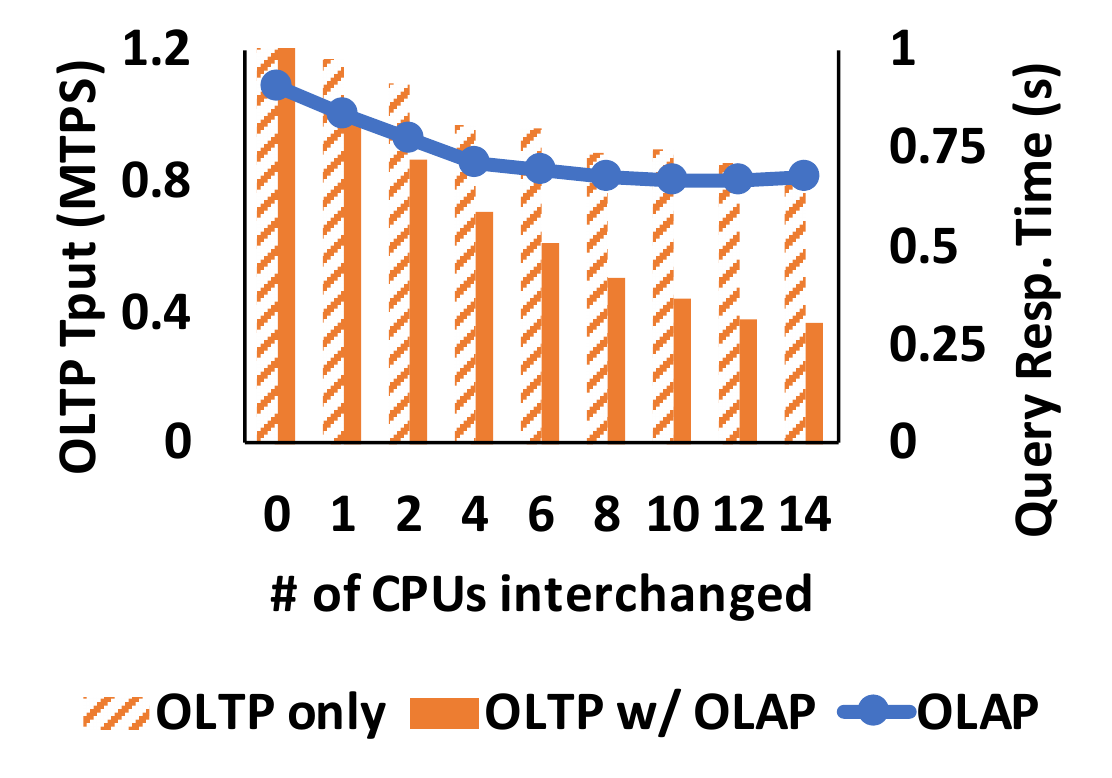} 
    \label{fig:colocated-oltp-olap}
  }%
  \end{minipage}%
  \hfill
  \begin{minipage}[t]{.33\textwidth}
  \centering
    \subfigure[OLTP/OLAP performance at State $S_2$]{\centering\includegraphics[width=\textwidth]{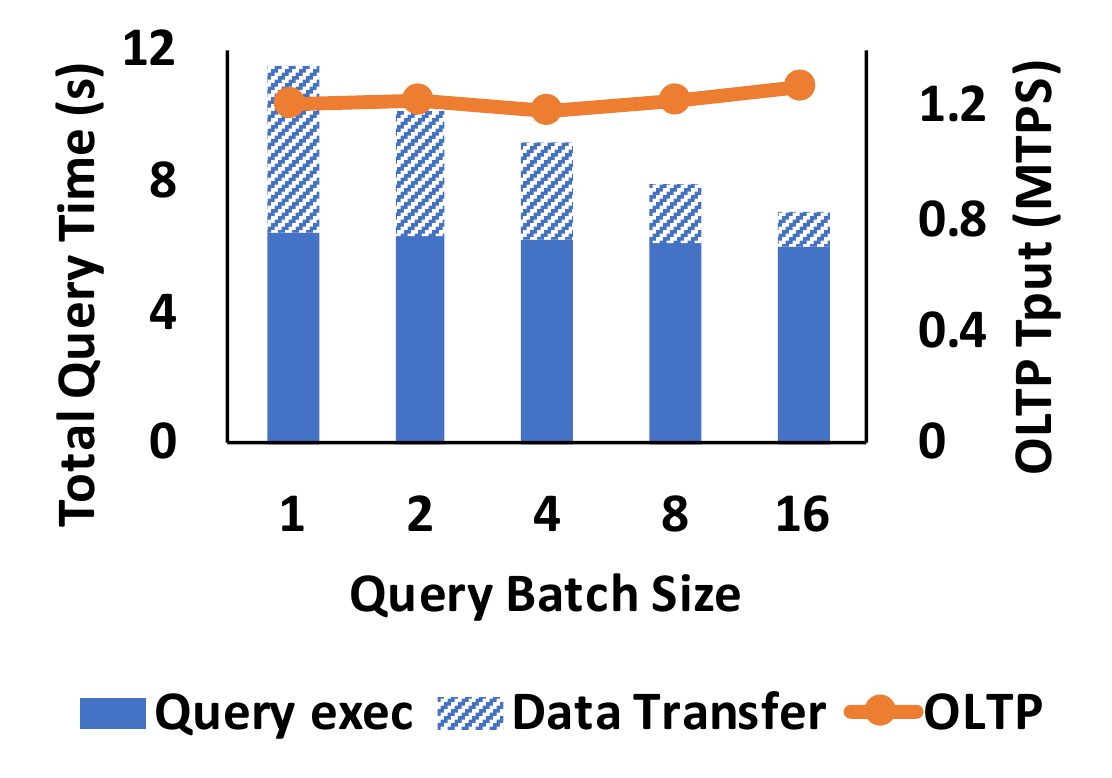} 
    \label{fig:scan_access_path}
    }%
  \end{minipage}%
  \hfill
  \begin{minipage}[t]{.33\textwidth}
  \centering
    \subfigure[OLTP/OLAP performance at State $S3_{NI}$]{\centering\includegraphics[width=\textwidth]{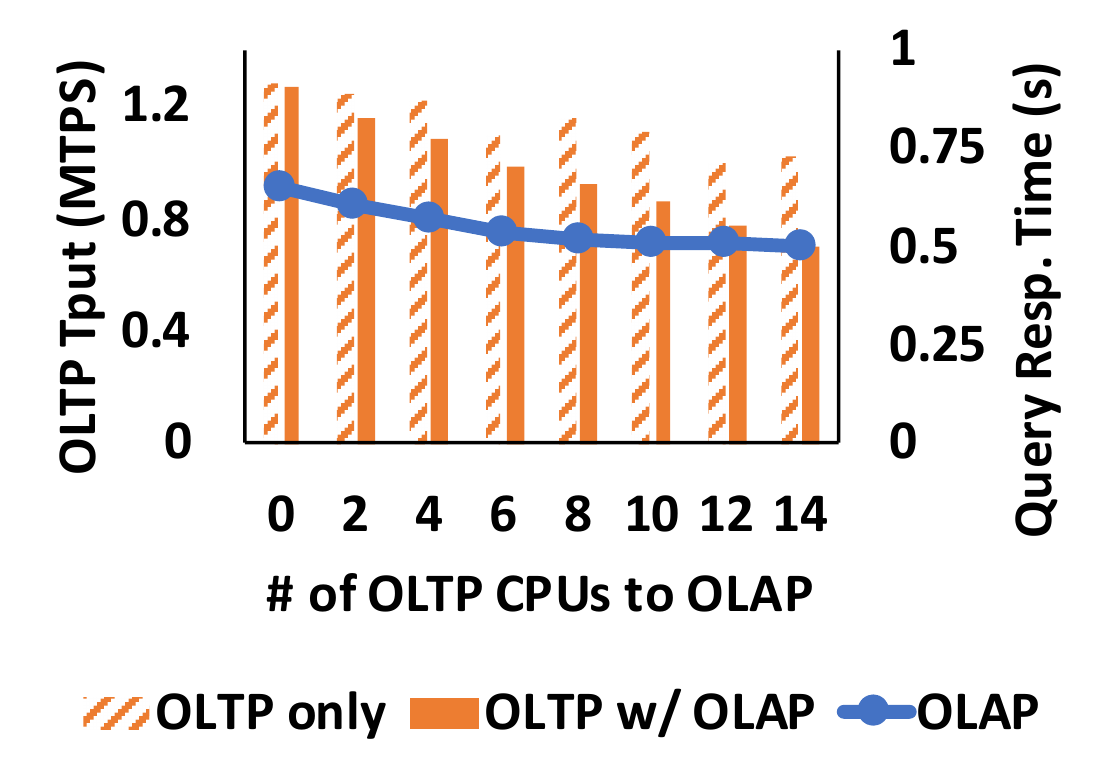} 
    \label{fig:resource_share}
    }%
  \end{minipage}%
  \caption
    {%
      Sensitivity analysis for states S1, S2, S3-NI
      \label{fig:sensitivityS1S2S3NI}%
    }%
\end{figure*}

\subsection{Hardware \& Software setup} \label{sec:eval_setup}
\textbf{Hardware.}
All the experiments were conducted on a server equipped with 2x14-core Intel Xeon Gold 6132 processor (32-KB L1I + 32-KB L1D cache, 1024-KB L2 cache, and 19.25-MB LLC) clocked at 2.60 GHz, with Hyper Threads, summing to a total of 56 hardware threads, and 1.5-TB of DRAM. 

\textbf{Software.}
Our RDE engine relies on 2-MB huge pages, and it pre-faults the memory before passing it to the OLTP and the OLAP engines to avoid any artifacts of memory allocations in the experimental results.
At system bootstrap, the OLTP and the OLAP engines get one CPU socket of the server, which corresponds to the full isolation state ($S_2$).
The OLTP engine uses the memory given by the RDE engine to create the two instances, its delta storage, its index and the rest of the data structures required.
Similarly, the OLAP engine uses the memory granted by the RDE engine to create the OLAP instance and initialize its buffers that will be needed for query execution.
Before every experiment, we execute a warm-up phase, and then we report steady-state analytical query response time and transactional throughput.


\textbf{Benchmark.}
We performed our experiments with the CH-benchmark~\cite{DBLP:conf/sigmod/ColeFGGKKKNNPSSSW11} which combines two industry standard benchmarks, TPC-C and TPC-H, for transactional and analytical processing systems, respectively. 
The schema inherits the relations specified in TPC-C and adds three more relations specified in TPC-H, which are \texttt{Supplier, Nation} and \texttt{Region}. 
To better analyze the effects of different HTAP schedules, we scale the database size following the TPC-H approach by a scale factor $SF$ and the size of the \texttt{LineItem} table becomes $SF*6,001,215$.
We fix 15 \texttt{OrderLines} per \texttt{Order} when initializing the database and we scale the number of records in \texttt{OrderLine} to $SF*6,001,215$. 
In contrast to TPC-H, and as per TPC-C specification, every \texttt{NewOrder} transaction generates five to fifteen order lines per order. 
Unless stated otherwise, all experiments are conducted on initial database with scale factor 300.
For the transactional workload, we assign one warehouse to every worker thread, which generates and executes transactions simulating complete transactional queue.
As the CH benchmark does not specify selectivities for conditions on dates, we consider 100\% selectivity, which is the worst case for join and groupby operations.
\revision{
Nevertheless, the selectivities do not affect the performance of our scheduling algorithm, since it assumes that the OLAP engine will perform full column scans in any case, as we explain in Section~\ref{sec:design-olap}.
Therefore, the amount of fresh data accessed does not depend on the selectivity.
}

\input{eval_sensitivity_rev.tex}

\input{eval_adaptive_scheduling.tex}

%% file: eval_sensitivity_rev.tex
\subsection{HTAP sensitivity analysis} \label{sec:eval_sensitivity_analysis}

In this section, we analyze the design states of our HTAP system and their impact on the performance of the OLTP and the OLAP engine.
We use Q1 and Q6, which perform aggregate operations over the \texttt{OrderLine} table, for two reasons: (i) they are simple queries that perform a scan over a single table, and therefore they can reveal overheads and opportunities, and, (ii) the \texttt{OrderLine} table grows in time, as the OLTP engine executes the \texttt{NewOrder} transaction.
Thus, we study the impact of data freshness-rate on the performance of both engines, at the different states that our system migrates.
We report the OLAP performance with query response time and the OLTP performance with throughput measured in million transactions per second (MTPS).

\textbf{Co-located OLTP and OLAP $\boldsymbol{[S_1]}$.}
In this state, we focus on the impact of resource allocation in the performance of each engine, due to hardware-level interference which is caused due to sharing CPU caches as well as memory and interconnect bandwidth.
We execute an experiment which considers that the engines are initially in full isolation (State $S_2$), and then they gradually trade CPUs from their sockets until they go half the way to each socket.
Both engines access the memory allocated by the OLTP engine, even though a different instance.
Therefore, initially the OLAP engine access all records from the remote socket, and it gradually gets local access as it trades its CPUs with the OLTP engine.
Instead, the OLTP engine initially has local access to its data, and gradually mixes it with remote accesses.

Figure~\ref{fig:colocated-oltp-olap} shows the performance of the OLTP and the OLAP engine.
The x-axis shows the number of CPU cores traded between the engines.
The y-axis on the left-hand side of the figure shows the performance of the OLTP engine (higher is better).
The y-axis on the right-hand side of the figure shows the query response time (lower is better).
We use striped bars to represent the transactional throughput when OLTP runs without any interference from OLAP and filled bars when OLAP and OLTP are executed concurrently to depict the effect of analytical query execution to the performance of transaction execution for all configurations.
We use a line to represent the query response time.
For every configuration that we report, we run a batch of experiments where Q6 is executed 16 times, one after the other, on the freshest snapshot of data, and we report the average performance, thereby considering that the amount of fresh data in the system increases.

As shown in the figure, the OLTP throughput drops up to 37\% in the absence of OLAP workload execution, but after the first 4 CPUs, the rate is smaller and almost stabilizes.
This happens because the transaction execution is dominated by random memory accesses and, therefore, the interconnect does not impose much overhead.
A part of this overhead is also accounted to cross-socket atomics, as has been reported in the OLTP literature~\cite{DBLP:journals/pvldb/AppuswamyAPIA17}.
The OLTP throughput drops up to 55\% in the presence of OLAP workload execution, whereas the rate of performance degradation is almost proportional for each set of CPUs that we trade.
This is accounted to the stress caused to the memory and the interconnect bandwidth by the OLAP query, which is a scan and dominates all memory accesses.
As the OLTP moves gradually to the remote socket, it performs more accesses through the CPU interconnect which is saturated by the OLAP engine, and therefore its performance drops following this pattern.
Accordingly, we observe that the OLTP engine throughput drops about 20\% due to the interference caused by OLAP workload execution.

Similar to the OLTP throughput, we observe that the OLAP response time improves up to the point where we trade 4 CPUs, where it starts stabilizing.
The reason is that the engine efficiently load balances across the two different interconnects, memory bus and remote CPU, and pipelines execution in such a way that it does not need further data locality.
Therefore, after 4 CPUs, the OLAP performance, while the OLTP performance keeps decreasing.

\textbf{Insight $\boldsymbol{[S_1]}$.}
The co-location of OLTP and OLAP engine creates interference which mostly affects the performance of the OLTP engine.
Furthermore, passing many data-local CPUs to the OLAP engine does not improve its performance, whereas it continues to hurt the performance of the OLTP engine.
Therefore, a bad the decision of the topology of each engine may hurt performance with no benefit.

\textbf{Isolated OLTP and OLAP $\boldsymbol{[S_2]}$.} 
In this state, we focus on the time required to copy the data between the engines, and how fast this cost can be amortized, by executing several queries over the same data on the OLAP engine.
We execute Q6 on the OLAP engine in different batches over the same snapshot of the data.
We vary the batch size from to 1 to 16 and we transfer the fresh data from the OLTP engine before the query execution starts.
We execute the same number of queries, regardless the batch size to make a fair comparison.

Figure~\ref{fig:scan_access_path} depicts the performance of the OLTP and the OLAP engine as we increase the batch size.
The x-axis shows the number of queries contained in each batch.
The y-axis on the left-hand side of the figure shows the cumulative query execution time in seconds (lower is better),
and on the right-hand side the throughput of the OLTP engine (higher is better).
We use solid bars to represent the response time \revision{for 16 query executions and stripped bars to represent the time spent on transferring data between the engines.}

As shown in the figure, the data transfer time almost equals the query execution time.
\revision{
Approximately 500MB of data are copied for each batch while 160MB are accessed by each query.
However, as we increase the batch size, the cost for copying the data gets amortized.
}
The OLTP throughput remains unaffected due to the physical isolation of the engines at the socket boundary.
The interference at the memory bus when the RDE engine copies the data to the OLAP socket does not affect the execution of transactions, since the OLTP engine does not fully utilize the memory bandwidth.

\textbf{Insight $\boldsymbol{[S_2]}$.}
Copying data from one engine to the other gets amortized after some time, provided that several queries access the same data.
Moreover, by periodically copying data to the OLAP engine, the system limits the stress on the OLTP memory bus, which is affected both when engines share their socket and when the OLAP engine reads the fresh data that it needs from the OLTP socket.
Therefore, $S_2$ brings the system into a steady state and, for this reason, the scheduler invokes it when the amount of fresh data becomes large enough.
\begin{figure}
  \centering
  \includegraphics[width=\columnwidth,height=4cm]{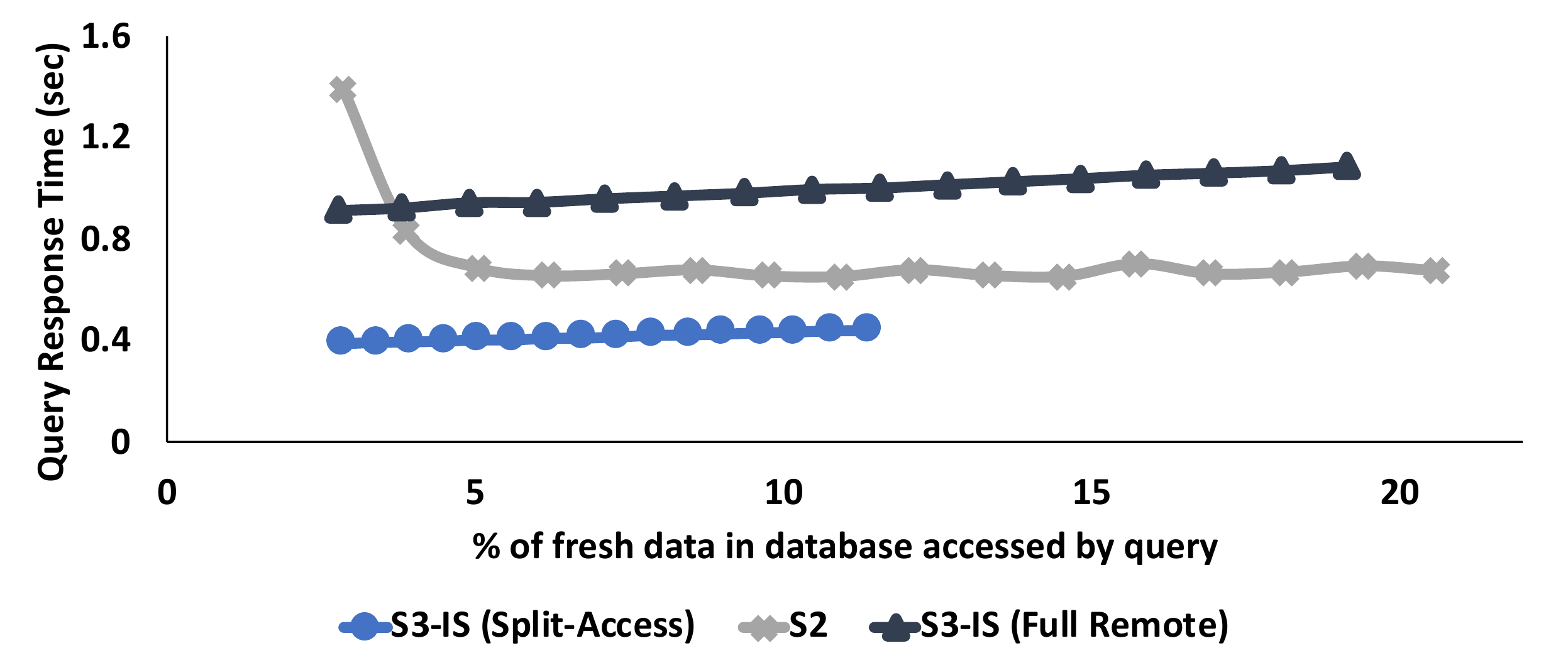}
  \caption
    {%
      OLAP response time with respect to data freshness\label{fig:data_pattern}%
    }%
\end{figure}
\textbf{Hybrid OLTP and OLAP $\boldsymbol{[S_3]}$.}
In this state, we focus on the impact of assigning OLTP CPU cores to the OLAP engine for the non-isolated case, which we refer to as $S_3-NI$ and of remote reads for the isolated case, which we refer to as $S_3-IS$.
Note that in $S3-IS$ we cannot store the data that we bring, because this can lead to an inconsistent snapshot on the OLAP side if some records are left behind because they are not needed by the current query.

Upon state migration, the OLTP engine switches instance and synchronizes the two instances.
At that point, the scheduler knows how many fresh records exist in the system from the RDE engine.
We use this information to optimize the access method of the OLAP engine enabling it to access from the OLTP engine only the data that are explicitly inserted.
We use this \textit{split-access} optimization only when the query accesses tables where data are \textit{inserted} and not \textit{updated}, as the latter would lead to accessing an inconsistent snapshot.

\begin{figure*}
  \centering
  \begin{minipage}[t]{0.5\textwidth}
  \centering
  \subfigure[OLAP]{\centering\includegraphics[width=\textwidth]{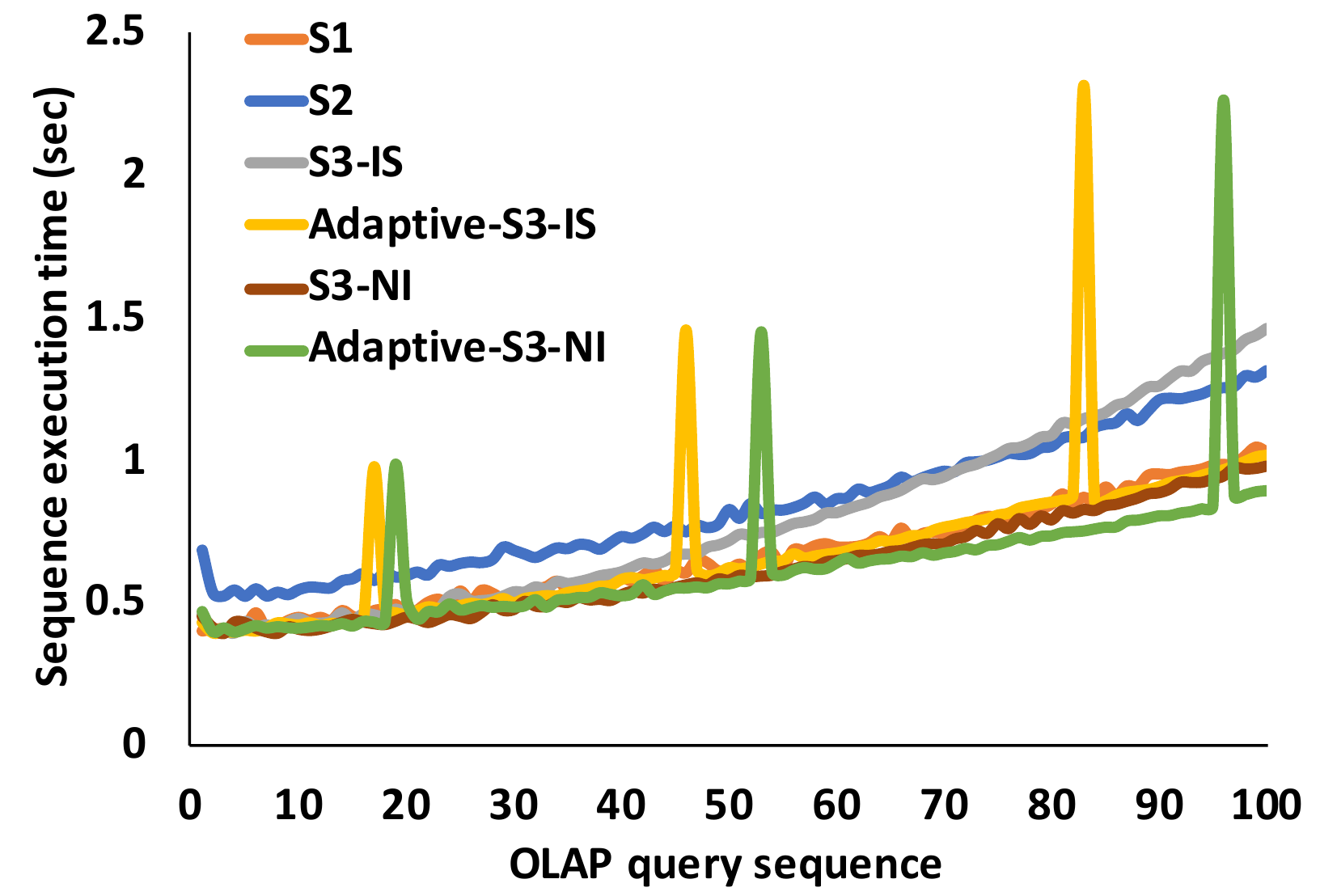} 
    \label{fig:seq-olap}
  }%
  \end{minipage}%
  \hfill
  \begin{minipage}[t]{0.5\textwidth}
  \centering
    \subfigure[OLTP]{\centering\includegraphics[width=\textwidth]{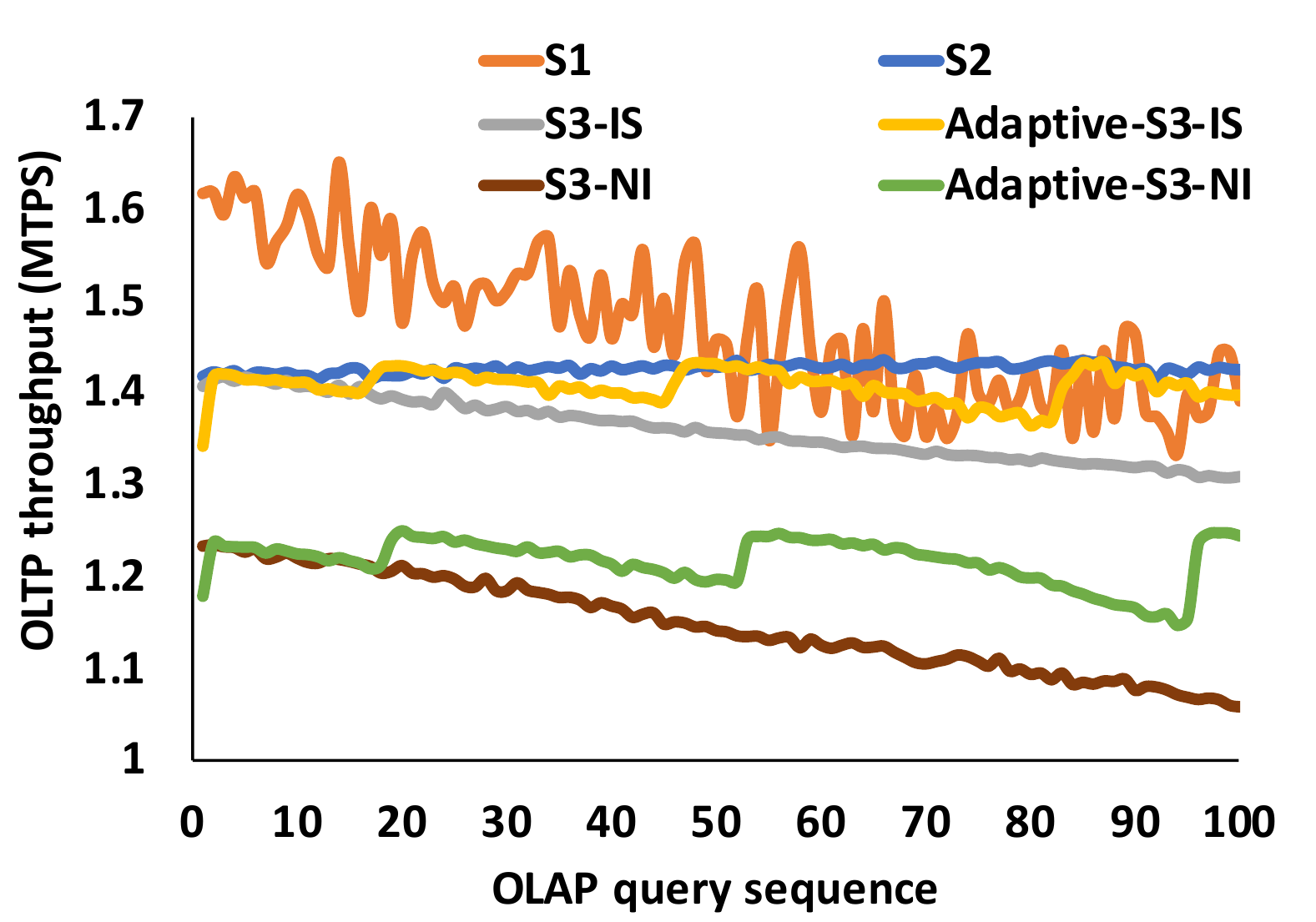} 
    \label{fig:seq-oltp}
    }%
  \end{minipage}%
  \hfill
  \caption
    {%
      HTAP performance under different scheduling states
      \label{fig:seq-adp}%
    }%
\end{figure*}

For $S_3-NI$, we vary the amount of CPUs that are passed from the OLTP to the OLAP engine.
We use Q1 for this case because Q6 introduces overheads that we explain in the following section and does not improve OLAP performance.
We use the split-access method when accessing the data, since this improves the execution time.
We report the results of our experiment in Figure~\ref{fig:resource_share} which has layout similar to Figure~\ref{fig:colocated-oltp-olap}.
As shown in the figure, the OLTP engine suffers from hardware-level interference, similar to the co-located case where we executed Q6.
Again, we observe that the performance of OLAP plateaus as we give it more that 6 CPUs, since it can already saturate the memory bandwidth.
The improvement that we are getting in the query response time is around 20\% in this experiment, because most of the data are accessed from the OLAP instance.
Effectively, this threshold corresponds to the maximum number of CPUs that the database administrator would allow the OLAP engine to elastically expand, while sacrificing OLTP performance.

For $S3-IS$, we execute an experiment where we vary the amount fresh data that the OLAP engine will need to access in order to achieve freshness-rate equal to 1.
\revision{In Figure~\ref{fig:data_pattern} we report the average query response time of Q1 when varying the percentage of fresh data accessed by the query. More specifically, we vary the fresh bytes in columns touched by the query and report in the x-axis the touched fresh bytes as a ratio over the total fresh bytes in the database.} 
The ratio of fresh data increases over time, as transactions are inserting new data.
There are two important observations to be made from this plot:
(i) The full-remote method for $S_3-IS$ is worse than $S_2$, since it always has to bring the same data over the interconnect.
$S_2$ is worse in the beginning because it needs to fetch more data for the first query, but as we execute the query many times, it stabilizes;
(ii) The split-access method for $S_3$ is better because it accesses only the fresh data that are required by the query.
Moreover, we observe that the blue line of the split-access $S_3$ approaches the grey line of $S_2$, and at some point they are expected to cross.
This point is when the scheduler prefers to do the data transfer before executing the query, because most of the data is fresh and the full ETL cost will be amortized.

\textbf{Insight $\boldsymbol{[S_3]}$.}
We have shown that an HTAP system benefits from transferring the data from the OLTP to the OLAP side, after some freshness threshold.
Moreover, hybrid execution provides benefits to query execution, as long as there are enough data to saturate the memory bandwidth throughout the whole query execution.
Finally, we have shown the amount of resources to be reallocated between the engines depends on the queries and the available hardware.

\revision{
\textbf{OLTP tail latency.}
As OLAP stresses the memory bus, the OLTP engine is expected to experience higher tail latencies. 
In $S_3-IS$ and $S_2$, this effect is expected to be smaller, as the OLAP accesses go through the CPU interconnect. 
However, this becomes higher as system migrates to $S_3-NI$, and to $S_1$ which is the worst case.
}

%% file: eval_adaptive_scheduling.tex
\subsection{Adaptive HTAP scheduling}\label{sec:expr_adaptive_htap}

In Figure~\ref{fig:seq-adp}, we evaluate the adaptive scheduling algorithm in HTAP and compare it with every individual state. 
We initialize the database at SF 30 before we synchronize the storage of both engines, thereby setting the freshness-rate of the OLAP instance to be 1.
The OLAP engine executes a set of 3 queries for 100 times in a sequence, one after the other and we report the total execution time of sequence, including any snapshotting or ETL.
We use two scan-heavy queries, Q1 \& Q6, and one join-heavy query, Q19. 
The OLTP engine executes TPC-C \texttt{NewOrder}, concurrently to the OLAP queries.
We remove the \texttt{LIKE} condition from Q19, as it is not supported by the OLAP engine.
In the following, we first describe the characteristics of every query and its interaction with the hardware, and then we evaluate their performance in the query mix.

\textit{CH-Q6 (scan-filter-reduce)} is memory intensive and has performance dependent on availability of total bandwidth that can be used to access data. In state S2, CH-Q6 executes at memory-bandwidth but pays an upfront cost of ETL operation. In state S1 and S3-NI, availability of more data-local bandwidth gives CH-Q6 a performance boost. In state S1, CH-Q6 can starve OLTP of memory-bandwidth due to long-sequential memory-scans and in reverse, can have interference with OLTP engine, as both, DRAM and interconnect bandwidth are shared with concurrent OLTP, stressing the DMA controllers of all NUMA sockets.

\textit{CH-Q1 (scan-filter-groupby)} is similar to Q6, but the final grouping and aggregation stress CPU-caches, and if executed across socket, based on group distribution, may get a performance hit by cache-coherency across NUMA nodes in a hash-based group-by operation. For less amount of fresh-data, CH-Q1 performs best with state S3-IS, consuming memory-bandwidth accumulative of interconnect bandwidth, followed by socket-local group-by. As the amount of fresh data increases, state S3-NI dominates in performance, hiding cross-socket load balancing effects.

\textit{CH-Q19 (fact-dimension-join)} joins a fact table with a dimension table, and is dominated by memory latency of random accesses, during the probing phase of hash-based join. The OLAP engine opts for broadcast-based join for CH-Q19 as the build side is relatively small (\textit{tpcc\_items} with 100,000 records) which penalizes cross-NUMA states, S1 and S3-NI. In addition to broadcast overhead, S1 faces additional interference with OLTP traffic across socket interconnect. However, with more ratio of fresh data, the broadcast cost is amortized by data-local memory accesses in state S3-NI.

\textbf{OLAP performance.} Figure~\ref{fig:seq-olap} shows the OLAP sequence execution times for different states. 
As the OLTP engine inserts data, execution time increases.
The selection of state for $S_3-IS$, $S_1$, and $S_3-NI$ depends on the performance requirements for the OLTP engine, from stricter to looser ones, respectively.
With the availability of elastic resources, the scheduler either opts for $S_3-NI$ or $S_1$, depending on the OLTP engine requirements.
Observe that the design of our OLTP engine with two instances, synchronized upon query arrival, inherently corresponds to the decision of our scheduler by setting $\alpha=0$, in the $S_1$ case.
As this approach for the OLTP engine provides flexibility to our design, we only report the performance for the adaptive $S_1$.

State $S_2$ is the slowest one, as it has to do an ETL for every query.
Given that we are not executing batches, the ETL cost only gets amortized with respect to $S3_{IS}$ after 75 query sets because at that point, in $S_2$ the OLAP engine has more socket-local data than in $S_3-IS$ where data are read from the remote socket.
State $S_3-NI$ provides further performance improvement due to presences of data-local compute resources where OLAP reduces fresh-data opportunistically and transfer less data over interconnect.
In all cases, we observe that their adaptive counterpart provides better performance, at the cost of a single ETL where one query pays with additional latency.
The timing of this ETL depends on the value of $\alpha$, which we currently set to 0.5.
Smaller values of $\alpha$ cause smaller tail latency, but at the cost of smaller benefit for the rest of the queries.
Finally, in all cases, we observe that the gap between the adaptive and the non-adaptive case is widening.
Across states, this gap goes up to 50\% ($Adaptive-S_3-NI$ and $S_3-IS$) showing the benefits to migrate from one state to an adaptive one.
We have executed further experiments with up to 300 queries of the same mix for the state $S_3-NI$ which has the slowest convergence, and we observed that this gap starts from 11\% in the given sequence, goes to 22\% for 200 queries, 25\% for 250 and 30\% for 300 queries.

\textbf{OLTP Performance.} 
Figure~\ref{fig:seq-oltp} shows the transactional throughput corresponding to OLAP query-sequence execution, under different system scheduling states.
OLTP throughput slightly degrades due to increased memory-pressure on OLTP-local DRAM as OLAP has hybrid-scans, that is, accessing data over interconnect in additional to OLAP-local scan. 
In all the adaptive schedules, OLTP throughput increases after every ETL operation which reduces pressure on OLTP-local memory-bus by OLAP, which corroborates the above claim.
S3-NI has a lower throughput compared to isolated counterparts due to the reallocation of CPU cores to the OLAP engine. 
Finally, $S_1$ has variance due to the co-location of an OLTP and and an OLAP in the same sockets.

\textbf{Insights.} Freshness-driven scheduling in HTAP adapts across feasible states, where feasibility is set as the workloads' isolation level. Isolated adaptive mode achieves 30\% speedup over S3-IS with 100 sequences. with 4-elastic cores, adaptive mode achieves 11\%, 22\% and 26\% performance gains at 100th, 200th and 250th sequence execution, and with time, adaptive schedule amortizes cost of data movement. In general, adaptive case builds upon hybrid-states while stressing to trigger ETL based on the freshness ratio in order to balance data-access across OLAP-local and fresh-data.

%% file: conclusion.tex
\section{Conclusion}
We look at HTAP as a scheduling problem, where the system balances OLTP and OLAP engine performance, depending on the data freshness and the performance requirements of the workload.
We define the HTAP design space ranging from fully co-located engines to fully isolated, and we devise an elastic system design which traverses across this space by distributing computing and memory resources to the OLTP and the OLAP engine.
We provide a scheduling algorithm which drives resource allocation decisions.
We perform a sensitivity analysis of our system, showing that exchanging resources between the two engines is beneficial until a certain point.
Finally, we evaluate the performance of our system using the CH-Benchmark, and we show that our system adapts to the data freshness and performance requirements of the workload.